\documentclass[showpacs,nofootinbib,showkeys,eqsecnum,prd,aps,preprint]{revtex4}

\usepackage[english]{babel}
\usepackage[cp1251]{inputenc}
\usepackage[T1]{fontenc}

\usepackage{amssymb}
\usepackage{amsmath}
\usepackage{amsfonts}
\usepackage{latexsym}
\usepackage{mathrsfs}
\usepackage{bm}
\usepackage{tensor}
\usepackage{hyperref}

\begin{document}

\title{Calculation of the wave functions of a quantum asymmetric top using the noncommutative integration method}

\author{A.I. Breev$^{1}$\thanks{
		breev@mail.tsu.ru}, and D.M. Gitman$^{3,2,1}$
	\thanks{
		dmitrygitman@hotmail.com} \\
	%EndAName
	{\normalsize $^{1}$ Department of Physics, Tomsk State University, Tomsk
		634050, Russia.}\\
	{\normalsize $^{2}$ P.N. Lebedev Physical Institute, 53 Leninsky prospekt,
		119991 Moscow, Russia;}\\
	{\normalsize $^{3}$ Institute of Physics, University of Sao Paulo, CEP
		05508-090, Sao Paulo, SP, Brazil; }
}

\begin{abstract}
	In this work, using the noncommutative integration method of linear 
	differential equations, we obtain a complete set of solutions to the 
	Schrodinger equation for a quantum asymmetric top in Euler angles. It is 
	shown that the noncommutative reduction of the Schrodinger equation leads 
	to the Lame equation. The resulting set of solutions is determined by the 
	Lame polynomials in a complex parameter, which is related to the geometry 
	of the orbits of the coadjoint representation of the rotation group. The 
	spectrum of an asymmetric top is obtained from the condition that the 
	solutions are invariant with respect to a special irreducible 
	$\lambda$-representation of the rotation group.
\end{abstract}



\maketitle

\section{Introduction}

The problem of the asymmetric top in quantum mechanics is a standard one, 
dealt with by a number of authors \cite{Landay, Ball, Zare, Wang1929, Casimir,Mull1941,King1943,Winter1954}. The quantum asymmetric top has 
many applications, ranging from quantum information \cite{Alb2020} to 
high-resolution spectroscopy \cite{Domingos2018}, especially in the fields of 
molecular \cite{Dennison1930,Lauro2020} and nuclear 
\cite{Bohr1953,Bohr1969,Cast1988} physics. It should be noted that the 
problem of scalar waves in a frozen Mixmaster universe is mathematically 
identical to that of the asymmetric top, 
except that half-integral angular momenta \cite{Hu1973a,Hu1974,Hu1973b}.

The stationary Schrodinger equation for a quantum asymmetric top does not 
allow separation of variables in the general case \cite{Landay}, because the 
necessary and sufficient conditions for the theorem on the separation of 
variables are not satisfied \cite{shth}. However, for 
wave functions that depend on only two variables, the equation admits a 
separation of variables in an elliptic coordinate system, which makes it 
possible to find the spectrum of an asymmetric top \cite{Sm1970}. In this 
case, the eigenfunctions in the standard approach are sought in the form of a 
series of Wigner $D$-functions. 

In this paper, to integrate the Schrodinger equation for a quantum 
asymmetric top, the noncommutative integration method of linear differential 
equations \cite{shap1995,sh2000,Br2020,mag2021} is used, which, unlike the 
method of separation of variables, allows us to reduce to an ordinary 
differential equation in the general case. Note that \cite{ShSO3} used this 
method to study the semiclassical spectrum of an asymmetric top.

The paper is organized as follows. In section \ref{S1} we introduce the basic 
concepts from the quantum theory of an asymmetric top and describe it in 
terms of the rotation group $SO(3)$. Section \ref{S2} is devoted to finding a 
complete set of solutions to the Schrodinger equation using the 
noncommutative integration method. In section \ref{S3}, we study the 
connection between the Wigner $D$-function and a special irreducible 
representation (irreps) of the group $SO(3)$, from which the completeness of 
the obtained set of solutions follows. In the last section \ref{S4}, we 
summarize and discuss the main results. Some useful technical details are 
placed in the Appendix.

Here we are using the natural system of units $c=\hbar=1$. 

\section{The quantum asymmetric top\label{S1}}

The Hamiltonian of the quantum top is defined as the operator
\begin{equation}
	\hat{H}=A\hat{L}_{1}^{2}+B\hat{L}_{2}^{2}+C\hat{L}_{3}^{2},\quad 
	A\geq B\geq C > 0,\label{opH}
\end{equation}
where $(2A)^{-1},(2B)^{-1}$and $(2C)^{-1}$ are the principal moments
of inertia, $\hat{L}_{a}$ are the components of the angular momentum
operator along the three principal axes of inertia of the top and
satisfying the commutation relations
\[
	[\hat{L}_{a},\hat{L}_{b}]=
	\hat{L}_{a}\hat{L}_{b}-\hat{L}_{b}\hat{L}_{a}
	=-i\epsilon_{abc}\hat{L_{c}},
\]
where $\epsilon_{abc}$ is the completely antisymmetric tensor with
$\epsilon_{123}=1$. The quantum spherical top corresponds to the
case $A=B=C$ and the quantum symmetrical top to the case $A=B\neq C$.
A top is called an asymmetric top if $A\neq B\neq C$. 

The square of the total angular momentum operator 
$\hat{L}^{2}=\hat{L}_{1}^{2}+\hat{L}_{2}^{2}+\hat{L}_{3}^{2}$
commutes with the Hamiltonian $\hat{H}$, $[\hat{L}^{2},\hat{H}]=0$. 

Let's describe quantum top problem in terms of rotation
group $SO(3)$. We describe the orientation of the top by Euler angles
$g=(\phi,\theta,\psi)$, $\phi\in[0;2\pi)$, $\theta\in[0;\pi)$,
$\psi\in[0;2\pi)$, referred to axes fixed in space. We note that
Euler angles parameterize the group element $g\in SO(3)$ (see Appendix).
Projections of the angular momentum operator $\hat{L}$ with respect
to the axes of the body-fixed reference frame are
\[
	\hat{L}_{1}=i\xi_{1},\quad\hat{L}_{2}=i\xi_{2},\quad
	\hat{L}_{3}=i\xi_{3},\quad [\hat{L}_{1},\hat{L}_{2}]=-i\hat{L}_{3},
\]
where $\xi_{a}$ are left-invariant vector fields (\ref{xi1}) (see Ref. 
\cite{Hu1973a}). Operators
for the components of angular momentum along the axes fixed in space
(space-fixed reference frame) are given by: 
\[
	\hat{J}_{x}=i\eta_{1},\quad \hat{J}_{y}=i\eta_{2},\quad
	\hat{J}_{z}=i\eta_{3},\quad [\hat{J}_{x},\hat{J}_{y}]=-i\hat{J}_{z},
\]
where $\eta_{a}$ are right-invariant vector fields (\ref{eta1}). The square of the total angular momentum operator is Casimir operator
of $SO(3)$ group and the same in both reference frames: 
\begin{eqnarray*}
\hat{L}^{2} & =K(-i\xi)=K(i\eta)=\\
 & -\frac{1}{\sin^{2}\theta}\left(\frac{\partial^{2}}{\partial\psi^{2}}+\frac{\partial^{2}}{\partial\phi^{2}}-2\cos\theta\frac{\partial^{2}}{\partial\phi\partial\psi}\right)
  -\frac{\partial^{2}}{\partial\theta^{2}}-\cot\theta\frac{\partial}{\partial\theta},
\end{eqnarray*}
where $K(f)=f_{1}^{2}+f_{2}^{2}+f_{3}^{2}$. 

There are three mutually commuting operators 
$\mathcal{X=}\{\hat{L}_{3},\hat{J}_{3},\hat{L}^{2}\}$.
We denote the common eigenfunctions of the set as 
$\left|j,m,n\right\rangle$,
\begin{eqnarray}
  \hat{L}^{2}\left|j,m,n\right\rangle &=& j(j+1)\left|j,m,n\right\rangle , \quad j=1,2,3,\dots,\nonumber \\
  -\hat{L}_{3}\left|j,m,n\right\rangle &=& n\left|j,m,n\right\rangle,
      \qquad\quad\,\,\,\, n=-j,\dots,j,\nonumber \\
  \hat{J}_{z}\left|j,m,n\right\rangle &=& m\left|j,m,n\right\rangle,
  \label{sysDv}
 \end{eqnarray}
They correspond to states with a given angular momentum $j$ and its
$z$-projection $n$ with respect to the axes of
the body-fixed reference frame and $z$-projection $m$ with respect
to the space-fixed reference frame. Explicit form
of the states $\left|j,m,n\right\rangle $ is given by the Wigner
$D$-functions that are matrix elements of the irreps of the group
$SO(3)$ (see Ref. \cite{Zare,Edmonds}):
\begin{eqnarray}
\langle g\mid j,m,n\rangle=D_{mn}^{j}(g)=e^{im\phi+in\psi}d_{mn}^{j}(\theta),\nonumber \\
d_{mn}^{j}(\theta)=(-1)^{m-n}\sqrt{\frac{(j+m)!(j-m)!}{(j+n)!(j-n)!}}\nonumber \\
\qquad\qquad\qquad\qquad\times\sin^{m-n}\frac{\theta}{2}\cos^{m+n}\frac{\theta}{2}P_{j-m}^{(m-n,m+n)}(\cos\theta),
\label{Dvigner}
\end{eqnarray}
where $P_{n}^{(\alpha,\beta)}(z)$ are Jacobi polynomials,
\begin{eqnarray*}
P_{n}^{(\alpha,\beta)}(z) & =\frac{(-1)^{n}}{2^{n}n!}(1-z)^{-\alpha}(1+z)^{-\beta}\frac{d^{n}}{dz^{n}}\left[(1-z)^{n+\alpha}(1+z)^{n+\beta}\right].
\end{eqnarray*}
The completeness and orthogonality conditions for the Wigner $D$-function
have the form:
\begin{eqnarray}
\frac{1}{8\pi^{2}}\int_{0}^{2\pi}d\psi\int_{0}^{\pi}\sin\theta d\theta\int_{0}^{2\pi}d\phi\overline{D_{mn}^{j}(g)}D_{\tilde{m}\tilde{n}}^{\tilde{j}}(g)=\frac{\delta_{j\tilde{j}}}{2j+1}\delta_{m\tilde{m}}\delta_{n\tilde{n}},\label{Dvig1}\\
\sum_{n=-j}^{j}\overline{D_{mn}^{j}(g)}D_{\tilde{m}n}^{j}(g)=\delta_{m\tilde{m}}.
	\label{Dvig2}
\end{eqnarray}

Thus, the Hamiltonian (\ref{opH}) is expressed in terms of the 
left-invariant vector fields on the group $SO(3)$, 
\[
	\hat{H}=H(-i\xi),\quad
	H(f)=Af_{1}^{2}+Bf_{2}^{2}+Cf_{3}^{2}
\]
and commutes with right-invariant vector fields $\eta_{a}$.
The states of a quantum top are determined by the wave function, which
is a function on the group $SO(3)$. 

There is a set of two mutually commuting symmetry operators
(the Casimir operator $K(-i\xi)$ and one of the operators $\xi_{a}$). The
state of the top with a certain value of quantum numbers $j$
and $m$ is described by the system
\begin{eqnarray}
 & \hat{H}\mid j,m\rangle=E\mid j,m\rangle,\label{sch}\\
 & \hat{L}^{2}\mid j,m\rangle=j(j+1)\mid j,m\rangle,\nonumber \\
 & \hat{J}_{z}\mid j,m\rangle=m\mid j,m\rangle.\nonumber 
\end{eqnarray}
Note that the system (\ref{sch}) cannot be solved by the separation method
variables. In the standart approach \cite{Landay,Hu1973a}, the solution of Eq. (\ref{sch})
is sought as: 
\begin{equation}
\mid j,m\rangle=\sum_{n=-j}^{j}a_{j,n}\left|j,m,n\right\rangle .
\label{expD}
\end{equation}
Substituting (\ref{expD}) in Eq. (\ref{sch}), we obtain a $(2j+1)$-
dimensional linear system 
\begin{equation}
\sum_{n=-j}^{j}a_{j,n'}\left(H_{n,n'}-E\delta_{n,n'}\right)=0,\quad H_{n,n'}=\langle j,m,n\mid\hat{H}\mid j,m,n'\rangle,
\label{sysC}
\end{equation}
 for the quantities $a_{j,n}$. The roots of equation
\begin{equation}
	\left\Vert H_{n,n'}-E\delta_{n,n'}\right\Vert =0
	\label{secEq}
\end{equation}
determine the energy levels of the top, after which the system of equations 
(\ref{sysC}) allows one to determine the wave functions of an asymmetric top 
with given values $j$ and $m$. 

The secular equation (\ref{secEq}) is of degree $2j+1$. In fact,
the eigenvalues $E$ are independent of $m$. This is so because the
Hamiltonian $\hat{H}$ commutes with the raising and lowering operators
$\hat{J}_{\pm}=\hat{J}_{1}\pm i\hat{J}_{2}$. The specialization of
the eigenvalue (\ref{sch}) equation to $m=0$,
\begin{eqnarray}
 & \hat{H}\mid j,0\rangle=E\mid j,0\rangle,\nonumber \\
 & \hat{L}^{2}\mid j,0\rangle=j(j+1)\mid j,0\rangle,\nonumber \\
 & \hat{J}_{z}\mid j,0\rangle=0.
 \label{schM0}
\end{eqnarray}
admits separation of variables in the elliptic coordinates \cite{Sm1970}. 
The elliptic coordinates $(\rho_{1},\rho_{2})$, $\rho_{1}\in(B,A)$,
$\rho_{2}\in(C,B)$, defined as
\begin{eqnarray*}
	\sin\theta\sin\psi & =\sqrt{\frac{(A-\rho_{1})(A-\rho_{2})}{(A-B)(A-C)}},\\
	\sin\theta\cos\psi & =\sqrt{\frac{(B-\rho_{1})(B-\rho_{2})}{(A-B)(A-C)}},
	\quad
	\cos\theta = \sqrt{\frac{(C-\rho_{1})(C-\rho_{2})}{(C-A)(C-B)}}.
\end{eqnarray*}
The solution of the system (\ref{schM0}) in elliptic coordinates 
$\psi_{j}(\rho_{1},\rho_{2})=\langle\rho_{1},\rho_{2}\mid j,0\rangle$
has the form:
\[
\psi_{j}(\rho_{1},\rho_{2})=\Lambda_{j}(\rho_{1})\Lambda_{j}(\rho_{1}),
\]
where function $\Lambda_{j}(\rho)$ satisfy the differential equation
\begin{equation}
\left\{ 4\sqrt{P(\rho)}\frac{d}{d\rho}\left(4\sqrt{P(\rho)}\frac{d}{d\rho}\right)-j(j+1)\rho+E\right\} \Lambda_{j}(\rho)=0,
\label{lameEq}
\end{equation}
which can readily be identified to be Lame equation in algebraic form. 

It follows from the theory of the Lame differential equation that for 
integers $j$ there are $2j+1$ linearly independent and mutually orthogonal 
functions $\Lambda_{j,s}(\rho)$ (the Lame polynomials) corresponding to 
$2j+1$ different eigenvalues $E_{j,s}$, $s=-j,\dots,j$. Solutions to the 
equation (\ref{lameEq}) are represented as series
\begin{eqnarray}
\Lambda_{j}^{(1)}(\rho)= & \sum_{k=0}^{\infty}a_{k}(\rho-B)^{j/2-k},\nonumber \\
\Lambda_{j}^{(2)}(\rho)= & \sqrt{\rho-A}\sum_{k=0}^{\infty}b_{k}(\rho-B)^{(j-1)/2-k},\nonumber \\
\Lambda_{j}^{(3)}(\rho)= & \sqrt{\rho-C}\sum_{k=0}^{\infty}c_{k}(\rho-B)^{(j-1)/2-k},\nonumber \\
\Lambda_{j}^{(4)}(\rho)= & \sqrt{(\rho-A)(\rho-C)}\sum_{k=0}^{\infty}d_{k}(\rho-B)^{j/2-k-1}.
\label{seriesLame}
\end{eqnarray}
From Eq. (\ref{lameEq}) the recurrent relations for the coefficients $a_{k}$, 
$b_{k}$, $c_{k}$ and $d_{k}$ follow, where $a_{l}=b_{l }=c_{l}=d_{l}=0$ for 
$l<0$. The eigenvalues of $E_{j,s}$ are obtained from the finiteness 
conditions for the series (\ref{seriesLame}):
\begin{equation}
a_{\left\lfloor j/2+1\right\rfloor }=b_{\left\lceil j/2\right\rceil }=c_{\left\lceil j/2\right\rceil }=d_{\left\lfloor j/2\right\rfloor }=0,
\label{condW}
\end{equation}
where $\left\lfloor x\right\rfloor =\max\{m\in\mathbb{N}\mid m\leq x\}$
is the floor function and $\left\lceil x\right\rceil =\min\{m\in\mathbb{N}\mid m\geq x\}$
is the ceiling function.

From (\ref{condW}) the eigenvalues $E_{j,s}$ are determined. For even values 
of $j$ we have one equation of degree $j/2+1$ and three equations of degree 
$j/2$ whose solution is $2j+1$ different eigenvalues $E_{j,s}.$ For odd 
values of $j$, we obtain three equations of degree $(j+1)/2$ and one equation 
of degree $j/2$, the solution of which is also $2j+1$ different eigenvalues 
$E_{j,s}$.

\section{The noncommutative integration\label{S2}}

In this section, we obtain a complete set of solutions to the stationary 
Schrodinger equation
\begin{equation}
	\hat{H}\Psi(g)=E\Psi(g).
	\label{kg}
\end{equation}  
with the Hamiltonian (\ref{opH}). Eq. (\ref{kg}) can be thought of as the 
quantum equation on the $SO(3)$ group. To construct solutions of quantum 
equations on the Lie groups, it is efficient to use the noncommutative 
integration method \cite{shap1995,sh2000,Br2020,ShSO3,br2019}. We apply 
this method to the Eq. (\ref{kg}).

First we must construct a special representation of the 
Lie algebra $\mathfrak{so}(3)$ of the group $SO(3)$. Following the papers 
\cite{shap1995,Br2020,ShSO3,br2019,br2014,br2016}, we introduce an 
irreducible $\lambda$-representation ($\lambda$-irrep) of the Lie algebra 
$\mathfrak{so}(3)$, which is parametrized by the parameter $j=1,2,\dots$ 
and acts in the space $\mathscr{ F}^{j}$ functions of the form
\begin{equation}
\Psi(q)=\sum_{n=-j}^{j}c_{n}e^{inq},\quad q\in Q,\quad c_{n}=\mathrm{const},\label{1.1}
\end{equation}
where $Q=\{q=\alpha+i\beta\mid\alpha\in[0;2\pi),\,\beta\in(-\infty,+
\infty)\}$. The $\lambda$-irrep of the Lie algebra 
$\mathfrak{so}(3)$ in space $\mathscr{F}^{j}$ is given by the operators 
\begin{eqnarray}
 & \ell_{1}(q,\partial_{q},j)=-i\sin q\partial_{q}+ij\cos q,\nonumber \\
 & \ell_{2}(q,\partial_{q},j)=-i\cos q\partial_{q}-ij\sin q,\nonumber \\
 & \ell_{3}(q,\partial_{q},j)=\partial_{q},\quad
   [\ell_a(q,\partial_{q},j),\ell_b(q,\partial_{q},j)]=\epsilon_{abc}\ell_c(q,\partial_{q},j).
   \label{lpr}
\end{eqnarray}
The operators $\hat{f}_{a}=-i\ell_{a}(q,\partial_{q},j)$ are Hermitian 
with respect to the scalar product
\begin{eqnarray}
 & (\Psi_{1},\Psi_{2})_{Q}^{j}=\int\overline{\Psi_{1}(q)}\Psi_{2}(q)d\mu_{j}(q),\quad\Psi_{1},\Psi_{2}\in\mathscr{F}^{j},\nonumber \\
 & d\mu_{j}(q)=C_{j}\frac{dq\wedge d\overline{q}}{\left(1+\cos(q-\overline{q})\right)^{j+1}},\quad C_{j}=\frac{(2j+1)!}{2^{j}(j!)^{2}},\label{scp}
\end{eqnarray}
and satisfy the commutation relations
\[
[\hat{f}_{a},\hat{f}_{b}]=i\epsilon_{abc}\hat{f}_{c},\quad K(\hat{f})=\hat{f}^2_1+\hat{f}^2_2+\hat{f}^2_3=j(j+1).
\]

The set of functions $\psi_{n}(q)=e^{inq}$ is orthogonal with respect to 
the scalar product (\ref{scp}): 
\begin{equation*}
	(\psi_{n},\psi_{\tilde{n}})_{Q}=\frac{1}{B_{nj}}\delta_{n\tilde{n}},
	\quad B_{nj}=\frac{(j!)^{2}}{(j-n)!(j+n)!}.
	%\label{psin23}
\end{equation*}
This follows the formula for the decomposition coefficients (\ref{1.1}):
\[
	c_{n}=B_{nj}(\Psi,\psi_{n})_{Q}.
\]
Then the generalized Dirac function in space $\mathscr{F}^{j}$,
\[
\Psi(q)=\int_{Q}\Psi(q')\delta_{j}(q,\overline{q'})d\mu_{j}(q'),\quad\Psi\in\mathscr{F}^{j}
\]
is given by the expression
\begin{equation}
	\delta_{j}(q,\overline{q'})=\sum_{n=-j}^{j}B_{nj}\psi_{n}(q)\overline{\psi_{n}(q')}=\frac{2j+1}{C_{j}}\left(1+\cos(q-\overline{q'})\right)^{j}.\label{defdel}
\end{equation}

The $\lambda$-irrep (\ref{lpr}) correspond to non-degenerate integer 
coadjoint orbit of the group $SO(3)$ passes through the covector 
$\lambda(j)=(j,0,0)$ (see Ref. \cite{sh2000}),
\begin{equation}
	\mathcal{O}_j = \{
		f\in\mathbb{R}^3\mid K(f)=j^2,\, f\neq 0
	\}.
	\label{orbit}
\end{equation}
The Kirillov form $\omega_j=(df_1\wedge df_2)/f_3$ in the orbit 
$\mathcal{O}_j$ sets a symplectic structure \cite{kirr1976}. It is well known that on the 
symplectic manifold, the Darboux canonical coordinates exist in which the 
symplectic form is canonical. There is the linear canonical transition 
\begin{eqnarray*}
	f_1(p,q,j) = -i p \sin q +j \cos q,\nonumber\\
	f_2(p,q,j) = -i p \cos q -j \sin q,\nonumber\\
	f_3(p,q,j) = p
	\label{canf}
\end{eqnarray*}
from the coordinates on the orbit to the Darboux coordinates 
$(p,q)$, $\omega_j = dp\wedge dq$. Note that the operators 
$\hat{f}_{a}$ can be considered as a result of 
$qp$-quantization of the orbit $\mathcal{O}_j$, $\hat{f}_{a} = f_a(\hat{p},\hat{q},j)$, $\hat{p} = -i\partial_q$, $\hat{q} = q$ (see Refs. \cite{ShSO3}), and 
the set $Q$ is a Lagrangian submanifold to the orbit $\mathcal{O}_j$. 

Within the noncommutative integration method of linear differential
equations, wave functions of asymmetric top are sought as a solution
system of equations
\begin{eqnarray}
 & \hat{H}(g)\Psi(q,j;g)=E\Psi(q,j;g),\label{Sch}\\
 & \left[\eta_{a}(g)+\ell_{a}(q,\partial_{q},j)\right]\Psi(q,j;g)=0,\label{XL}
\end{eqnarray}
where the operators $\eta_{a}$ commute with Hamiltonian, 
$[\hat{H}(g),\eta_{a}(g)]=0$.
Note that since
\[
[\eta_{a}(g)+\ell_{a}(q,\partial_{q},j),\eta_{b}(g)+\ell_{b}(q,\partial_{q},j)]=\epsilon_{abc}\left(\eta_{c}(g)+\ell_{c}(q,\partial_{q},j)\right),
\]
then the system (\ref{XL}) is consistent. Integration of the system of 
equations (\ref{XL}) gives
\begin{eqnarray}
\Psi(q,j;g) & =\left(\cos\theta+i\cos(q+\phi)\sin\theta\right)^{j}\label{seeksol}\\
 & \times\Phi_{j}\left(2\arctan\left[e^{i\theta}\cot\left(\frac{q+\phi}{2}\right)\right]\right).\nonumber 
\end{eqnarray}
Substituting (\ref{seeksol}) into the Eq. (\ref{Sch}), we obtain the 
reduced equation
\begin{equation}
	H(-i\ell(q',\partial_{q'},j))\Phi_{j}(q')=E_{j}\Phi_{j}(q').
	\label{redH}
\end{equation}
The Eq. (\ref{redH}) describes a quantum asymmetric top in 
$\lambda$-representation. The Hamiltonian 
$H(-i\ell(q',\partial_{q'},j))=H(\hat{f})$ can be considered as the result of 
$qp$-quantization of the classic top on the coadjoint orbit  
(\ref{orbit}).

Taking into account the explicit form of the $\lambda$-irrep 
operators (\ref{lpr}), we obtain an ordinary differential equation for 
the function $\Phi_{j}(q')$:
\begin{eqnarray}
	\left\{ (A\sin^{2}q'+B\cos^{2}q'-C)\frac{d^{2}}{dq'^{2}}+(\sin q'\cos q')(1-2j)(A-B)\frac{d}{dq'}\right.\label{redeq1}\\
	+\left.(A\cos^{2}q'+B\sin^{2}q')j^{2}+j(A\sin^{2}q'+B\cos^{2}q')-E_{j}\right\} \Phi_{j}(q')=0.\nonumber 
\end{eqnarray}

Remarkable is the fact that by replacing
\begin{eqnarray}
 & \Phi_{j}(q')=\left(\frac{2(A-C)(B-C)}{\rho(q')-C}\right){}^{j/2}\Lambda_{j}\left(\rho(q')\right),\label{PhiLame}\\
 & \rho(q')=\frac{2(A-C)(B-C)}{A+B-2C-(A-B)\cos2q'}+C,\nonumber 
\end{eqnarray}
 we get the Lame equation in algebraic form (\ref{lameEq}) on the function $\Lambda_{j}(\rho)$:
\begin{eqnarray*}
 & \left\{ 4\sqrt{P(\rho)}\frac{d}{d\rho}\left(4\sqrt{P(\rho)}\frac{d}{d\rho}\right)-j(j+1)\rho+E_j\right\} \Lambda_{j}(\rho)=0.
\end{eqnarray*}
Thus, we arrive at the same equation (\ref{lameEq}) as in the case of 
separation of variables in the elliptic coordinate system, but on a 
function of the complex variable $\rho=\rho(q')$. As will be shown below, 
in contrast to the set of solutions $\psi_{j}(\rho_{1},\rho_{2})$, the 
set of solutions $\Psi(q,j;g)$ forms a complete set.

Solutions which respond to Lame polynomials (\ref{seriesLame}) will be 
labeled as follows:
\[
\Phi_{j}^{(N)}(q')=\left(\frac{2(A-C)(B-C)}{\rho-C}\right)^{j/2}\Lambda_{j}^{(N)}(\rho),\quad\rho=\rho(q'),\quad N=1,2,3,4.
\]
Let's write it explicitly:
\begin{eqnarray*}
\Phi_{j}^{(1)}(q') & =\left(2(A-B)(B-C)\right)^{j/2}\\
 & \times\sum_{k=0}^{\infty}\frac{a_{k}}{(B-C)^{k}}\left(\cos q'\right)^{j-2k}\left(a-\cos^{2}q'\right)^{k},\\
\Phi_{j}^{(2)}(q') & =i\left(2(A-B)(B-C)\right)^{j/2}\sqrt{\frac{A-C}{B-C}}\sin q'\\
 & \times\sum_{k=0}^{\infty}\frac{b_{k}}{(B-C)^{k}}\left(\cos q'\right)^{j-2k-1}\left(a-\cos^{2}q'\right)^{k},\\
\Phi_{j}^{(3)}(q') & =\left(2(A-B)(B-C)\right)^{j/2}\sqrt{\frac{A-C}{A-B}}\\
 & \times\sum_{k=0}^{\infty}\frac{c_{k}}{(B-C)^{k}}\left(\cos q'\right)^{j-2k-1}\left(a-\cos^{2}q'\right)^{k},\\
\Phi_{j}^{(4)}(q') & =i\left(2(A-B)(B-C)\right)^{(j-1)/2}\sqrt{A-C}\sin q'\\
 & \sum_{k=0}^{\infty}\frac{d_{k}}{(B-C)^{k}}\left(\cos q'\right)^{j-2k-2}\left(a-\cos^{2}q'\right)^{k},
\end{eqnarray*}
where $a=(A-C)/(A-B)$. 

Functions $\Phi_{j}^{(N)}(q')$ belong to the function space 
$\mathscr{F}^{j}$ only when they contain positive powers in $\cos q'$. 
For this it is necessary and sufficient that
\begin{equation}
	a_{1+j/2}=b_{j/2}=c_{j/2}=d_{j/2}=0,
	\label{cond1}
\end{equation}
for even $j$ and
\begin{equation}
	a_{(j+1)/2}=b_{(j+1)/2}=c_{(j+1)/2}=d_{(j-1)/2}=0,
	\label{cond2}
\end{equation}
for odd values of $j$. Note that the conditions 
(\ref{cond1})--(\ref{cond2}) coincide with the conditions (\ref{condW}) 
and they determine the known spectrum $E_{j,s}$ of the quantum asymmetric 
top.

The coefficients $a_{0}$, $b_{0}$, $c_{0}$ and $d_{0}$ are defined
from the normalization condition for eigenfunctions
\[
	\Phi_{j,s}(q')=\left(\frac{2(A-C)(B-C)}{\rho-C}\right)^{j/2}\Lambda_{j,s}\left(\rho\right)
\]
of the operator $H(-i\ell(q',\partial_{q'},j))$ corresponding to the 
eigenvalues $E_{j,s}$:
\begin{equation}
	\left(\Phi_{j,s},\Phi_{j,s'}\right)_{Q} = (2 j+1)\delta_{ss'}.
	\label{normq}
\end{equation}
%
%. Вообще говоря, коэф-ты a0--d0 различны для различных s.
%
Then the wave functions 
\begin{eqnarray}
	\Psi_{q,j,s}(g) & =\langle g\mid q,j,s\rangle=\left(\cos\theta+i\cos(q+\phi)\sin\theta\right)^{j}\nonumber \\
 & \times\Phi_{j,s}\left(\psi+2\arctan\left[e^{i\theta}\cot\left(\frac{q+\phi}{2}\right)\right]\right)\label{mySolT}
\end{eqnarray}
corresponding to the eigenvalues $E_{j,s}$ satisfy the Eq. (\ref{Sch}) 
and the normalization condition
\[
	\langle q,j,s\mid q,j,s\rangle =  
	 \frac{1}{8\pi^{2}}\int_{0}^{2\pi}d\psi\int_{0}^{\pi}\sin\theta 
	 d\theta\int_{0}^{2\pi}d\phi\left|\Psi_{q,j,s}(g)\right|^{2}=  
	 \delta_{j}(q,\overline{q}).
\]

We give expressions for the wave functions (\ref{mySolT}) and the 
eigenvalues $E_{j,s}$ for the lowest values of the quantum number $j$. 
For $j=0$ we have $E_{0,1}=0$ and $\Psi_{q,0,1}=1$. For $j=1$ we have
\[
E_{1,1}=A+C,\quad E_{1,2}=A+B,\quad E_{1,3}=B+C
\]
 and wave functions
 \begin{eqnarray*}
	\Psi_{q,1,1}(g) &=&
	\sqrt{3}\bigg{\{}\frac{1}{2}\left[\cos(q-\theta+\phi)+cos(q+\theta+\phi) + 2i\sin\theta\right]\cos\psi\\
	&-&\sin(q+\phi)\sin\psi\bigg{\}},\\
\Psi_{q,1,2}(g) & =&\sqrt{3}\left(\cos\theta+i\cos(q+\phi)\sin\theta\right),\\
\Psi_{q,1,3}(g) & =&\sqrt{3}\left(\cos\psi\sin(q+\phi)+\left[\cos\theta\cos(q+\phi)+i\sin\theta\right]\sin\psi\right),
\end{eqnarray*}
respectively. 

Thus, we have obtained a set of solutions (\ref{mySolT}), which is 
parameterized by a set of parameters $\{q,j,s\}$, where $j$ is the 
quantum number corresponding to the eigenvalues operator $\hat{L}^2$, 
$q\in Q$ is a complex number that is not an eigenvalue of 
any integral of motion. 

\section{Wigner $D$-function and $\lambda$-representation of $\mathfrak{so}(3)$ group\label{S3}}

Note that the solution (\ref{seeksol}) can be represented in the integral
form
\begin{equation}
\Psi(q;g)=T^{j}(g)\Phi(q)=\int_{Q}\mathscr{D}_{qq'}^{j}(g)\Phi(q')d\mu(q'),\quad\Phi(q')\in\mathscr{F}^{j},\label{seeksol2}
\end{equation}
where the kernel
\begin{eqnarray}
\mathscr{D}_{qq'}^{j}(g) & =\left(\cos\theta+i\cos(q+\phi)\sin\theta\right)^{j}\nonumber\\
    &\qquad\qquad\qquad\times\delta_{j}\left(\psi+2\arctan\left[e^{i\theta}
    \cot\left(\frac{q+\phi}{2}\right)\right],\overline{q'}\right)
    \label{Du}
\end{eqnarray}
satisfies the system of equations
\begin{eqnarray}
 & \left(\eta_{a}(g)+\ell_{a}(q,\partial_{q},j)\right)\mathscr{D}_{qq'}^{j}(g)=0,\nonumber \\
 & \left(\xi_{a}(g)+\overline{\ell_{a}(q',\partial_{q'},j)}\right)\mathscr{D}_{qq'}^{j}(g)=0,\label{sysD}
\end{eqnarray}
with the initial condition $\mathscr{D}_{qq'}^{j}(0,0,0)=\delta_{j}(q,\overline{q'})$.
Using the Eq. (\ref{defdel}) for the generalized delta function, we get
\begin{eqnarray}
	\mathscr{D}_{qq'}^{j}(g) & =\frac{2^{j}(j!)^{2}}{(2j)!}
	\bigg{\{}
	[\cos(\phi+q)\cos(\overline{q'}-\psi)+1]\cos\theta\nonumber \\
 & +i\left[\cos(\phi+q)+\cos(\overline{q'}-\psi)\right]\sin\theta\nonumber \\
 & +\sin(\phi+q)\sin(\overline{q'}-\psi)\bigg{\}}^{j}.\label{fD}
\end{eqnarray}

It is shown in \cite{sh2000} that for a generalized function satisfying 
the system of equations (\ref{sysD}) on some unimodular Lie group, the 
relations
\begin{equation}
\mathscr{D}_{qq'}^{j}(g\cdot g')=\int_{Q}\mathscr{D}_{qq''}^{j}(g)\mathscr{D}_{q''q'}^{j}(g')d\mu_{j}(q''),\quad\overline{\mathscr{D}_{qq'}^{j}(g)}=\mathscr{D}_{q'q}^{j}(g^{-1}).\label{qqq}
\end{equation}
From Eq. (\ref{qqq}) implies that the operators $T^{j}(g)$ in 
(\ref{seeksol2}) are the operators of the unitary 
$\lambda$-irrep of the group $SO(3)$ in the space 
$\mathscr{F}^{j}$.
%Отметим, что представление 
%$T^{j}(g)$ представляет собой индуцированное действие группы:
%\[
%	T^{j}(g)\Phi(q)=U^{j}(q,g)\Phi(q\cdot g),
%	\quad U^{j}(q,g)=\left(\cos\theta+i\cos(q+\phi)\sin\theta\right)^{j},
%\]

Let us find a connection between the Wigner $D$-function (\ref{Dvigner}) 
and the $\mathscr{D}_{qq'}^{j}(g)$ function. Note that the 
Wigner $D$-function can be uniquely defined as a solution to the system 
of equations (\ref{sysDv}) with the initial condition 
$D^j_{m n}(0,0,0) = \delta_{m n}$. Then we will look for a solution to 
the system (\ref{sysDv}) in the form
\begin{equation}
D_{mn}^{j}(g)=C_{mn}^{j}\int_{Q}\overline{F_{m}(q)}\Phi_{n}(q')\mathscr{D}_{qq'}^{j}(g)d\mu_{j}(q)d\mu_{j}(q').\label{repDD}
\end{equation}
Substituting (\ref{repDD}) into (\ref{sysDv}), by the functions 
$F_{m}(q)$ and $\Phi_{n}(q')$ we get
\begin{eqnarray}
 & -i\ell_{3}(q,\partial_{q},j)F_{m}(q)=mF_{m}(q),\label{fff}\\
 & -i\ell_{3}(q',\partial_{q'},j)\Phi_{n}(q')=n\Phi_{n}(q').
\end{eqnarray}
Whence $F_{m}(q)=e^{imq}$, $\Phi_{n}(q')=e^{inq'}$. The coefficient 
$C_{mn}^{j}$ has the form
\[
	C_{mn}^{j}=\sqrt{B_{nj}B_{mj}}exp\left[\frac{i\pi}{2}(m-n)\right].
\]
From (\ref{Dvig1})--(\ref{Dvig2}) and the relation (\ref{repDD}) it 
follows that the functions $\mathscr{D}_{qq'}^{j}(g)$ are complete and 
orthogonal:
\begin{eqnarray}
 & \int\limits _{G}\overline{\mathscr{D}_{qq'}^{j}(g)}\,\mathscr{D}_{\tilde{q}\tilde{q}'}^{\tilde{j}}(g)d\mu(g)=\frac{1}{2j+1}\delta_{j}(q,\bar{\tilde{q}})\delta_{j}(q',\bar{\tilde{q}}')\delta_{j\tilde{j}},\nonumber \\
 & \sum_{j=0}^{\infty}(2j+1)\int\limits _{Q\times Q}\overline{\mathscr{D}_{qq'}^{j}(g)}\mathscr{D}_{qq'}^{j}(\tilde{g})d\mu_{j}(q)d\mu_{j}(q')=\delta(g\cdot\tilde{g}^{-1}).\label{fullD}
\end{eqnarray}
The expression (\ref{repDD}) can be written in the form
\begin{eqnarray*}
 	& \mid j,m,n\rangle=\sqrt{B_{nj}B_{mj}}e^{i\pi(m-n)/2}\int_{Q\times Q} 
 	e^{inq'-im\overline{q}}d\mu_{j}(q)d\mu_{j}(q')
 	\mid j,q,q'\rangle.
\end{eqnarray*}
where $\langle g\mid j,q,q'\rangle=\mathscr{D}_{qq'}^{j}(g)$. Here is an 
expression for the expansion of the states $\mid j,q,q'\rangle$ in terms 
of Wigner $D$-functions $\mid j,m,n\rangle$:
\begin{equation}
\mid j,q,q'\rangle=\sum_{n=-j}^{j}\sum_{m=-j}^{j}\sqrt{B_{nj}B_{mj}}e^{-in\overline{q'}+imq-i\frac{\pi}{2}(m-n)}\mid j,m,n\rangle.\label{invjqq}
\end{equation}

Since the operator $H(-i\ell(q',\partial_{q'},j))$ is Hermitian in the 
space $\mathscr{F}^j$ with respect to the inner product (\ref{scp}), then 
the set of eigenfunctions $\Psi_{q,j,s}(g)$ is complete:
\begin{equation}
	\sum_{s=-j}^j \frac{\overline{\Phi_{j,s}(q')}\Phi_{j,s}(q')}{2j+1} = 
	\delta_j(q,\overline{q'}),
	\label{full1}
\end{equation}
where the sum over $s$ means the sum over the spectrum of 
the asymmetric top. From (\ref{full1}) and (\ref{fullD}) follows the 
completeness of the set (\ref{mySolT}), 
\begin{eqnarray}
 & \mid q,j,s\rangle=\int_{Q}d\mu_{j}(q')\,\Phi_{j,s}(q')\mid j,q,q'\rangle,\label{SecSol}\quad
   \Psi_{q,j,s}(g)=\langle g\mid q,j,s\rangle,\nonumber\\
 & \sum_{j=0}^\infty
	\sum_{s=-j}^j
	\int_Q d\mu_j(q)
	\mid q,j,s\rangle \langle q,j,s \mid\, = 1.\nonumber
\end{eqnarray}

\section{Concluding remarks}\label{S4}

In this article, using the noncommutative integration method, we find the 
complete system of solutions to the Schrodinger equation (\ref{sch}) for 
an asymmetric top. It is shown that the noncommutative reduction of the 
Schrodinger equation reduces it to the Lame equation (\ref{lameEq}), 
which arises when separating variables in an elliptic coordinate system 
for the states $\mid j,0\rangle$. In this approach, the spectrum of the 
asymmetric top arises from the requirement that the solution of the 
reduced equation (\ref{redeq1}) belong to the space $\mathscr{F}^{j}$, 
which is invariant under the irreducible $\lambda$-irrep of the 
group $SO(3)$. In this case, the set of solutions is expressed by an 
explicit formula (\ref{mySolT}), in contrast to the solution in the form 
of a series (\ref{expD}). It is shown that the complete set of solutions 
is parametrized by the quantum number $q$, which takes complex values 
from the set $Q$, which has the Lagrangian submanifold to the coadjoint orbit 
of the $SO(3)$ group.

We also obtained a connection (\ref{repDD}) between the kernels 
$\lambda$-irrep and the Wigner $D$-function, which allowed us to 
show the completeness of the set (\ref{mySolT}).

Note that the constructed solutions are not coherent 
\cite{Malkin1979,Klauder1985,Perelomov1986,Gazeau2009}, 
since they do not minimize the uncertainty relation 
$(\Delta K)^{2}\geq j$. Since in states that corresponding to the basis 
$\mid q,j,s\rangle$ we have
\begin{eqnarray*}
	(\Delta K)^{2}&= & j(j+1)\delta_{j}(q,\overline{q})\\
	&=& \frac{2^{j}(j!)^{2}}{(2j)!}j(j+1)
	\left(\cosh2\mathrm{Im}q\right)^{j}\\
	&\geq & \frac{4^{j}(j!)^{2}}{(2j)!}j(j+1)\geq j^{2}\log4>j,
	\quad j>0.
\end{eqnarray*}
Therefore, the states $\mid q,j,s\rangle$ differ significantly from the 
coherent states of the quantum asymmetric top \cite{shelepin}.

By a direct check, one can verify that the set of states 
\begin{eqnarray}
\mid j,m,s\rangle & =\sqrt{B_{mj}}\int_{Q}d\mu_{j}(q)\,e^{-im\overline{q}}\mid q,j,s\rangle\nonumber\\
 & =\sqrt{B_{mj}}\int_{Q\times Q}d\mu_{j}(q)d\mu_{j}(q')\,e^{-im\overline{q}}\Phi_{j,s}(q')\mid j,q,q'\rangle
 \label{mainSol}
\end{eqnarray}
are eigenstates for the complete set of operators 
$\{\hat{H},\hat{L}^{2},\hat{J}_{3}\}$ and satisfies Eq. (\ref{sch}) for 
$\mid j,m\rangle=\mid j,m,s\rangle$. Thus, the states of an asymmetric 
top with a given value of $j$ and $m$ are determined by the expression 
(\ref{mainSol}).

The sets of solutions (\ref{mySolT}) and (\ref{mainSol}) can be 
useful for studying the static mixmaster cosmological model (see Refs. 
\cite{Hu1973a,Hu1974,Hu1973b,prit1985,Dowker1974}).

\section*{Acknowledgements}

The work is supported by Russian Science Foundation, 
Grant No. 19-12-00042. Gitman is grateful to CNPq for continued support.

\appendix

\section{Vector fields on the group $SO(3)$\label{App}}

In the Lie algebra $\mathfrak{so}(3)$ of the rotation group $SO(3)$ we 
introduce some basis $\{e_{1},e_{2},e_{3}\}$, with respect to the commutation 
relations of the algebra have the form $[e_{a},e_{b}]=C_{ab}^{c}e_{c}$, where
the structure constant $C_{ab}^{c}=\epsilon_{abc}$ is the completely
antisymmetric tensor with $\epsilon_{123}=1$, $[\cdot,\cdot]$ denotes
the Lie brackets; $a,b,c=1,\dots,3$. The adjoint representation matrices
$(ad_{a})_{b}^{c}=[e_{a},e_{b}]^{c}$ have the form
\begin{eqnarray*}
\mathrm{ad}_{1}=
\left(\begin{array}{ccc}
0 & 0 & 0\\
0 & 0 & -1\\
0 & 1 & 0
\end{array}\right),\quad 
\mathrm{ad}_{2}=\left(\begin{array}{ccc}
0 & 0 & 1\\
0 & 0 & 0\\
-1 & 0 & 0
\end{array}\right),\quad 
\mathrm{ad}_{3}=\left(\begin{array}{ccc}
0 & -1 & 0\\
1 & 0 & 0\\
0 & 0 & 0
\end{array}\right).
\end{eqnarray*}
The group element of the $SO(3)$ will be parametrized using the Euler angles 
$\phi$, $\theta$, and $\psi$:
\[
g(\phi,\theta,\psi)=g_{z}(\varphi)g_{x}(\theta)g_{z}(\psi)\in SO(3),\quad\varphi,\psi\in[0;2\pi),\quad\theta\in[0;\pi),
\]
where $g_{x}(t)$, $g_{y}(t)$ and $g_{z}(t)$ are rotation matrices by the 
angle $t$ about the axes $Ox$, $Oy$ and $Oz$ respectively:
\begin{eqnarray}
g_{x}(t)&=&e^{t\mathrm{ad_{1}}}=\left(\begin{array}{ccc}
1 & 0 & 0\\
0 & \cos t & -\sin t\\
0 & \sin t & \cos t
\end{array}\right),\nonumber\\
g_{y}(t)&=&e^{t\mathrm{ad_{2}}}=\left(\begin{array}{ccc}
\cos t & 0 & \sin t\\
0 & 1 & 0\\
-\sin t & 0 & \cos t
\end{array}\right),\nonumber\\ 
g_{z}(t)&=&e^{t\mathrm{ad_{3}}}=\left(\begin{array}{ccc}
\cos t & -\sin t & 0\\
\sin t & \cos t & 0\\
0 & 0 & 1
\end{array}\right).\label{A1}
\end{eqnarray}
To find generators of an arbitrary irrep of $SO(3)$, one has to examine
representations in the space of functions $f=f(\varphi,\theta,\psi)$
on the group. The left regular representation $T^{L}(g)$ acts in
the space of functions $f(g)$, $g=g(\varphi,\theta,\psi)\in SO(3)$,
on the group as follows \cite{barut}:
\[
T^{L}(g')f(g)=f(g'^{-1}\cdot g),\quad g'\in SO(3),
\]
whereas the right regular representation $T^{R}(g)$ acts in the same
space as follows:
\[
T^{R}(g')f(g)=f(g\cdot g'),\quad g'\in SO(3).
\]
The decomposition of the left (and right) regular representation contains
any irrep of the group. 

For generators that correspond to the one-parameter subgroup $\omega(t)$
in the left $T^{L}(g)$ and right $T^{R}(g)$ regular representations,
we obtain, respectively: 
\[
\xi_{\omega(t)}f(g)=\left.\frac{d}{dt}T^{R}(\omega(t))f(g)\right|_{t=0},\quad\eta_{\omega(t)}f(g)=\left.\frac{d}{dt}T^{L}(\omega(t))f(g)\right|_{t=0}.
\]
Vector fields $\xi_{\omega(t)}$ are called left-invariant vector fields, and 
$\eta_{\omega(t)}$ are called right-invariant vector fields corresponding to 
the subgroup $\omega(t)$. Let us choose one-parameter
subgroups as (\ref{A1}). The straightforward calculations yield the
following expressions for generators \cite{shelepin}:
\begin{eqnarray}
 & \xi_{1}=\xi_{g_{x}(t)}=\frac{\sin\psi}{\sin\theta}\frac{\partial}{\partial\phi}+\cos\psi\frac{\partial}{\partial\theta}-\cot\theta\sin\psi\frac{\partial}{\partial\psi},\nonumber \\
 & \xi_{2}=\xi_{g_{y}(t)}=\frac{\cos\psi}{\sin\theta}\frac{\partial}{\partial\phi}-\sin\psi\frac{\partial}{\partial\theta}-\cot\theta\cos\psi\frac{\partial}{\partial\psi},\nonumber \\
 & \xi_{3}=\xi_{g_{z}(t)}=\frac{\partial}{\partial\psi}
 \label{xi1}
\end{eqnarray}
and
\begin{eqnarray}
\eta_{1} & =\eta_{g_{x}(t)}=\cot\theta\sin\phi\frac{\partial}{\partial\phi}-\cos\phi\frac{\partial}{\partial\theta}-\frac{\sin\phi}{\sin\theta}\frac{\partial}{\partial\psi},\nonumber\\
\eta_{2} & =\eta_{g_{y}(t)}=-\cot\theta\cos\phi\frac{\partial}{\partial\phi}-\sin\phi\frac{\partial}{\partial\theta}+\frac{\cos\phi}{\sin\theta}\frac{\partial}{\partial\psi},\nonumber\\
\eta_{3} & =\eta_{g_{z}(t)}=-\frac{\partial}{\partial\phi}.
\label{eta1}
\end{eqnarray}

The following standard commutation relations hold:
\[
[\xi_{a},\xi_{b}]=\epsilon_{abc}\xi_{c},\quad[\eta_{a},\eta_{b}]=\epsilon_{abc}\eta_{c},\quad[\xi_{a},\eta_{b}]=0.
\]

%\section*{References}

\end{document}